\documentclass[useAMS,usenatbib]{mn2e}
\usepackage{multirow}
\usepackage{graphicx}
\usepackage{footnote}
\usepackage{hyperref}
\usepackage{amsmath}
\usepackage{soul}
\usepackage[usenames,dvipsnames]{pstricks}
\graphicspath{{Images/}}
\title{Featureless Classification of Light Curves}
\author{S.D. K\"ugler, N. Gianniotis, K.L. Polsterer\\
Heidelberg Institute for Theoretical Studies, Schloss-Wolfsbrunnenweg 35,
69118 Heidelberg, Germany}
\begin{document}

\date{}

\pagerange{\pageref{firstpage}--\pageref{lastpage}} \pubyear{2002}

\maketitle

\label{firstpage}

\begin{abstract}
In the era of rapidly increasing amounts of time
series data, classification  of variable
objects has become the main objective of time-domain astronomy.
Classification of irregularly sampled time series is particularly
difficult because the data cannot be represented naturally as a 
vector
which can be directly fed into a classifier. In the literature,
various statistical features serve as vector representations. 

In this work, we represent time series by a density model.
The density model captures all the information available, including
measurement errors. Hence, we view this model as a generalisation to the static
features which directly can be derived, e.g., as moments from the density.
Similarity between each pair of time series is quantified by the distance
between their respective models. Classification is performed
on the obtained distance matrix.

In the numerical experiments, we use data from the OGLE and ASAS surveys and
demonstrate that the proposed representation performs up to par with the
best currently used feature-based approaches. The density representation
preserves all static information present in the observational data, in
contrast to a less complete description by features.
The density representation is an upper
boundary in terms of information made available to the classifier.
Consequently, the predictive power of the proposed classification depends
on the choice of similarity measure and classifier, only. Due to its principled
nature, we advocate that this new approach of representing time series has
potential in tasks beyond classification, e.g., unsupervised learning.
\end{abstract}


\section{Introduction}

The variation of the brightness of an astronomical object over time (hereafter
called light curve or time series) is an important way to obtain knowledge and
constraint properties of the observed source. With the advent of large sky
surveys such as the Large Synoptical Sky survey (LSST,
\citealp{2011arxiv...0805...2366I}) the incoming data stream will be so immense
that the applied methodology has to be reliable and fast at the same time.
While the origin of variability can be very different, a huge fraction of the
variable objects in the sky has a stellar origin. From those variable stars
many show (quasi-) periodic behaviour and originate from the instability stripe
in the Hertzsprung-Russell-diagram or are multi-star systems where the origin of
the variability is the mutual occultation. The main focus of this work will be
on periodic sources, but in principle the presented methodology can also
be used for non-periodic sources \citep[see e.g.][]{2013arxiv...1310...1976D}.

The classification performance of periodic sources is already fairly high
provided that the period and the amplitude of the variation are determined
correctly \citep{1899ApJ....10..255B, 1902AnHar..38....1B,refId0}. But apart
from the very soft boundaries between the classes, the quality of the
period-finding algorithm depends on the type of variability ifself
\citep{2013MNRAS.434.3423G} and thus a dependency between those two properties
is encountered. In order to break this dependency one can either rely on only
\mbox{(quasi-)} static features\footnote{Throughout this paper we will divide
features derived by other authors in three categories:
\textbf{non-static}: everything directly related to period finding and 
features derived from the periodogram. \textbf{quasi-static}: features 
that treat the data as function instead instead of a time series, e.g.
slope, linear trend. \textbf{static:} all features that treat the measured
fluxes only as an ensemble and thus the temporal information is discarded, e.g.
median, standard deviation. A complete list of features used here is given in
Table \ref{featTable}.} for the classification or estimate the period and
derive classifications by analysing the phase-folded light curves
\citep[see e.g.][and references herein]{2007AA...475.1159D}.
\citet{2011ApJ...733...10R} showed that the inclusion of static features yields
an improved classification performance and that the contribution of the static
and non-static features to the accuracy is of the same order.

In this work, we introduce a novel representation of time series that aims to
replace the static features. We represent each noisy data  
point by a Gaussian; the mean of the respective Gaussian is the measurement and
the standard deviation is given by the measurement uncertainty 
(photometric error). Hence, every time series is represented by a mixture of
Gaussians that conserves all static information available in the data. We
advocate that this a simple and natural choice. In contrast to that,
features can be seen as derivatives (such as moments) of this density model and
therefore only describe certain properties of it.
For instance, \citet{Moments} show that moments are just able to describe the
tails of a distribution but do not necessarily give a good description of the
underlying distribution. 

As a consequence, the proposed density-based
representation presents an upper boundary to the static information content
which can be made available to the classifier. The similarity of two densities
is thereby judged using three widely used distance measures, the $L2$-norm, the
Kullback-Leibler-divergence and the Bhattacharyya distance. These measures of
similarity are then fed into two different classifiers. Finally, we compared
the classification performance of the density- and feature-based approaches.

The aim of this work is to introduce an alternative and more general notion of
similarity between light curves, which correctly takes into account measurement
uncertainty. In the new representation all static information contained in the
observations are conserved in a more principled way and adjacently fed to
the classifier. Consequently, we expect that this new representation
provides a reference in terms of classification performance.

In Section \ref{sec:method}
the new representation and its respective application to the classifier are
described. After describing the used data in Section \ref{sec:data}, the
results of two different experiments are presented in Section \ref{sec:results}.
We conclude with a discussion of our approach in Section \ref{sec:discussion}.
%  the available
% information content. With our approach we are thus able to estimate the accuracy
% performance encoded in static features and have thus a reliable baseline for the performance of other classifiers including also other feature sets.
% Finally, we can now retain meaningful distances between two sets of time series
% without relying on an arbitrary feature choice. This allows for the application
% of clustering and visualization techniques in a meaningful way.
% 
% As a consequence, we hope that the concept of generalisation becomes a manifest
% in the time-domain community and that also generalised concepts for the non-static
% features can be developed in the near-future, making the search and definition
% of new features obsolete.

\section{Method}
\label{sec:method}
\begin{figure}
\includegraphics[width=\columnwidth]{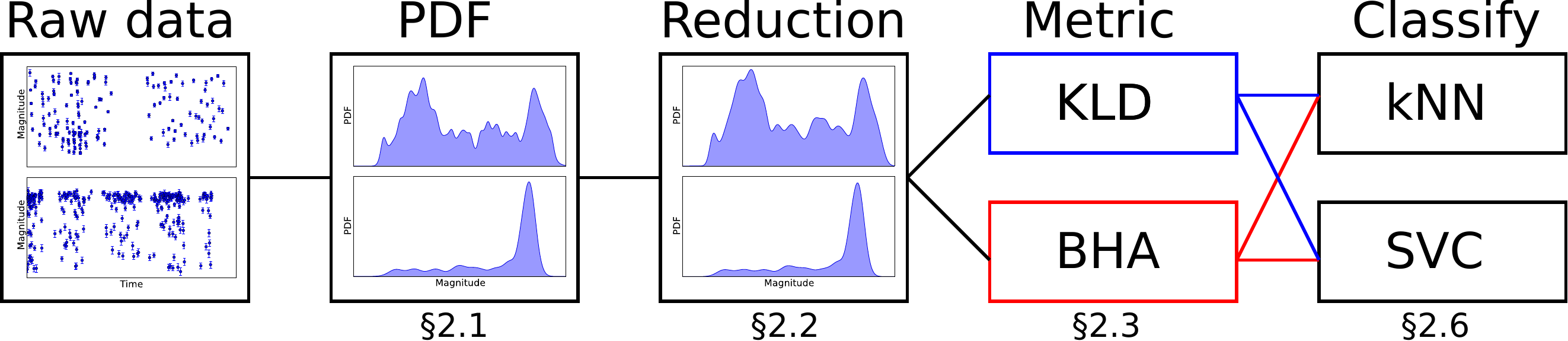}
\caption{\label{fig:scheme}Schematic view of all steps from the raw
data to the classification method.}
\end{figure}

In this section the methodology is described. A sketch of the entire
classification process is shown in Figure \ref{fig:scheme}. Each step is
annotated with the respective subsection in the text; the FCLC software
which includes all steps described in the following is available at 
\href{http://ascl.net/1505.014}{http://ascl.net/1505.014}.
\subsection{Converting data points into densities}
\begin{figure}
\centering
\includegraphics[width=0.49\textwidth]{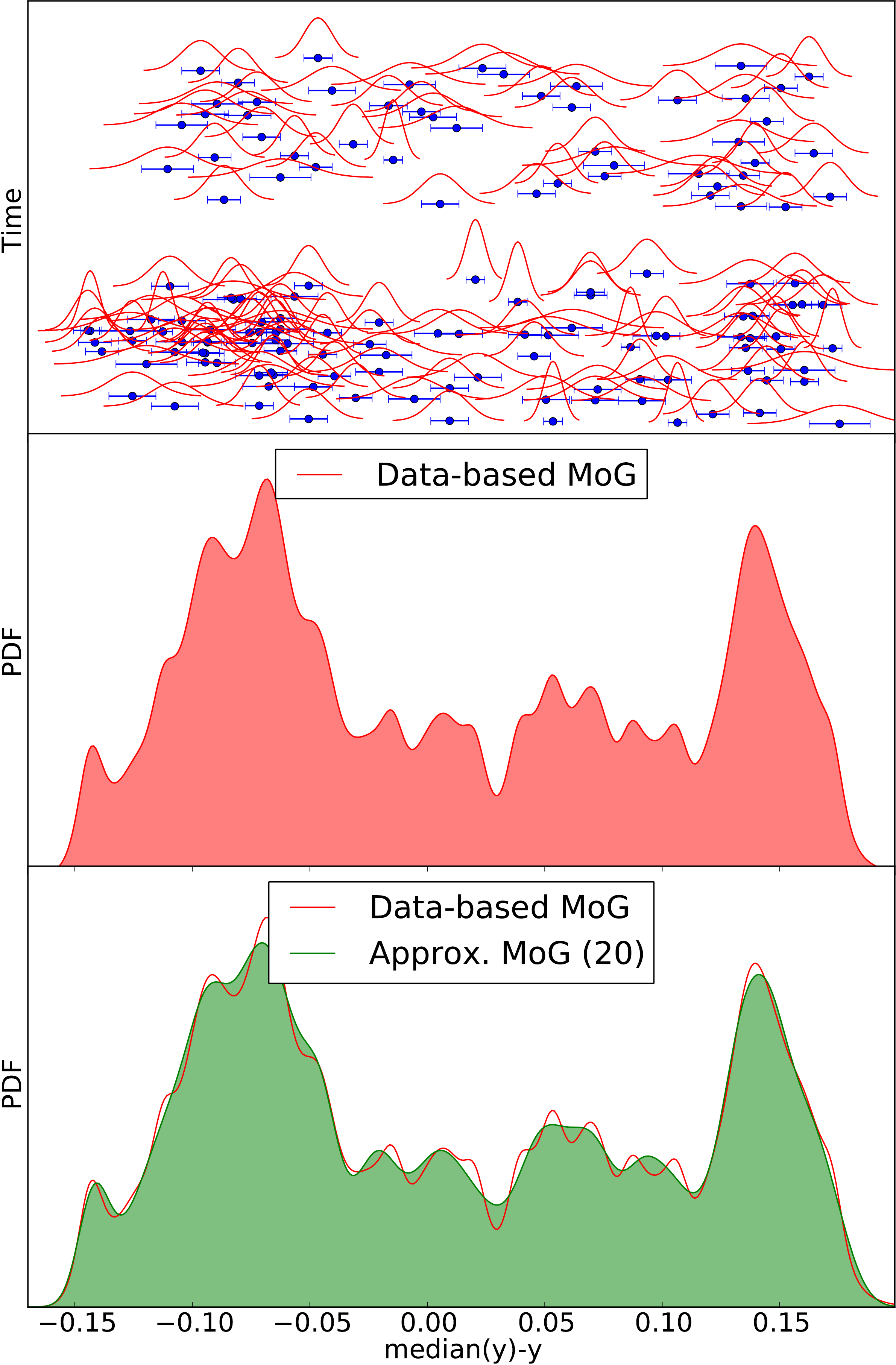}
\caption{The principle of the conversion to densities,
every point is described by a normal distribution which are then added up to a
PDF. \label{fig:hist2cont}}
\end{figure}

The key idea of our method is to convert the individual data points with their
errors as a continuous density. We treat each data point as a normal
distribution with a mean $\mu$ equal to the magnitude $y$ and a width $\sigma$
equal to the photometric error $\Delta y$ of the respective measurement.
This allows us to convert the discrete $M$ number of observations into a
continuous density by using:
\begin{equation}
PDF\left( x \right) =
\frac{1}{M}\sum_{i=1}^{M}\mathcal{N}\left(x \mid y_i,\Delta y_i\right) 
\end{equation}
where $\mathcal{N}\left(\mu, \sigma\right)$ is the
normal distribution with expectation $\mu$ and width $\sigma$, which returns
the probability of the occurrence for a given value $x$.
Each light curve is, after subtracting the median, converted to such a
probability density function ($PDF$);  a visualisation of this process is shown
in Figure \ref{fig:hist2cont}. This idea was already mentioned in the work of
\citet{Bhatta}.

% \textbf{We'd like to point out that the treatment of individual data points as
% densities (as well as for features) is only valid if the observations are 
% randomly drawn from the underlying distribution. That means that the number of
% draws (observations) has to be statistically significant and the observation
% times should not in any way be correlated with e.g., the period of the system.
% As fortunately the observations are a superposition of seasonal effect, weather
% impact and observation strategy, all observations of a single source can indeed
% be seen as ``uncorrelated''.}

\subsection{Parsimonious mixtures of Gaussians}
An important step to make the computation of the distances computationally
feasible is to reduce the number of Gaussians in the mixture of Gaussians
(MoG). The look-up function for individual values of $x$ scales linearly with
the number of Gaussians in the mixture. The computation of the distance
between every two densities scales directly with the number of Gaussians $m$
in each density. Thus the computational complexity for computing the
distance matrix is of the order of $\mathcal{O}\left(n^2m\right)$, where $n$
is the total number of light curves to be classified. This computation gains a
significant speedup by reducing the number of Gaussians; reducing the
number of observations (typically $M=300$) to a mixture of Gaussians with $m=20$
components yields an effective gain in speed of 
$\left(\frac{300}{20}\right)\approx15$.

We tested several ways described in the literature to reduce the number of
Gaussians effectively \citep{redMOG}. After experimenting with the different
methods we found that the method by \citet{Runnalls} yielded the most
satisfactory results. The basic idea is that two similar (in terms of
the Kullback-Leibler-Divergence, see below) Gaussians can be approximated by a
single normal distribution.
The dissimilarity between two normal distributions with amplitudes $W_0; W_1$
means $\mu_0; \mu_1$ and widths $\sigma_0; \sigma_1$ is thereby measured by
\begin{equation}
D = 0.5 \omega 
\mathrm{log}\left(
\tilde{\omega}_0\frac{{\sigma_0}^{2\tilde{\omega}_1}}{{\sigma_1}^{2\tilde{\omega}_1}}+
\tilde{\omega}_1\frac{{\sigma_1}^{2\tilde{\omega}_0}}{{\sigma_0}^{2\tilde{\omega}_0}}+
\tilde{\omega}_0\tilde{\omega}_1
\frac{\left(\mu_1-\mu_0\right)^2}{\sigma_0^{2\tilde{\omega}_0}
\sigma_1^{2\tilde{\omega}_1}}\right)
\end{equation}
with $\omega = \sigma_0 + \sigma_1$; $\tilde{\omega}_0 =
\frac{\sigma_0}{\omega}$; $\tilde{\omega}_1 = \frac{\sigma_1}{\omega}$. The pair
of normal distributions with the closest distance $D$ is then merged into a
new single Gaussian with weight $W_{01} = W_0+W_1$, expectation
$\mu_{01}=\frac{W_0}{W}\mu_0+\frac{W_1}{W}\mu_1$ and variance
$\sigma^2_{01}=\frac{W_0}{W}\sigma^2_0+\frac{W_1}{W}\sigma^2_1+
\frac{W_0W_1}{W^2}\left(\mu_0-\mu_1\right)^2$. The search and replacement is
then performed iteratively until the desired number of new components is
reached. An example of a reduced MoG is shown in the bottom plot in Figure
\ref{fig:hist2cont}.
Apart from the decreased computational complexity the reduction in number of
components used in the MoG has yet another very interesting side effect. Due to
the loss of information the new $PDF$ is always just a smoothed version of 
the density-based on the real data. As the data are irregularly sampled this
 smoothing is effectively a better representation of the true underlying
 density. 
 Obviously, the number of Gaussians to be used is a parameter which has to be
 optimised. Here, it will be optimised by maximizing the classification
 accuracy for a given dataset and classifier.
  
 Another aspect to mention is the conservation of outliers. Since iteratively
 only the most similar Gaussians are merged into a single one, the presence
 and probability of outliers will remain unchanged throughout this procedure.
%    s = np.sqrt(w0/W*s0**2 + w1/W*s1**2 + w0*w1/W**2 * (m0-m1)**2)

\subsection{Similarity of probability densities}
After converting all light curves to $PDF$, we apply different measures of
similarity between two given probability densities $P(x)$, $Q(x)$. As light
curves differ in apparent magnitude we subtract the median magnitude in order
to align the densities of different objects.
\subsubsection{$L2$-norm}
The most obvious choice for comparing two densities is the $L2$-norm,
defined as
\begin{equation}L2\left(P(x), Q(x)\right) 
%= \int \left(\tilde{P}(x)-\tilde{Q}(x)\right)^2 dx
= \int \left(P(x)-Q(x)\right)^2~dx~.\end{equation} 
While the $L2$-norm is a very robust and reliable measure of the similarity, it
is not very sensitive to faint tails as differences in the main component are
penalized more heavily. But, as stated in the introduction, the tails contain
the vast majority of information of a density. Hence, we do not expect the
$L2$-norm to be a good distance measure for our classification problem.

\subsubsection{Bhattacharyya distance}
The Bhattacharyya distance ($BHA$), defined as,
\begin{equation}BHA\left(P(x), Q(x)\right)
%-\mathrm{log}\left(\int \sqrt{P(x)Q(x)}dx \right)
= -\mathrm{log}\int
\sqrt{P(x)Q(x)}dx\end{equation}
is a generalisation of the Mahalanobis distance which, in contrast to the latter
one, takes into account the difference in shape. The Bhattacharyya distance has
been used in classification problems before \citep[see e.g.,][]{Bhatta} and
thus seems a very good choice for our method.

\subsubsection{Symmetrised Kullback-Leibler divergence}
The Kullback-Leibler divergence ($KLD$), 
\begin{equation}KLD\left(P(x), Q(x)\right) %= \int
%\left(P(x)\mathrm{log}\left(\frac{P(x)}{Q(x)}\right)\right)dx 
= \int P(x)\mathrm{log}\left(\frac{P(x)}{Q(x)}\right)dx
\end{equation}
is a measure of similarity of two probability densities in information
theory. It consists of two terms one being the entropy (information content) of
$P(x)$ and a term which is the expectation of
$\mathrm{log}\left(Q(x)\right)$ with respect to $P(x)$. The second term is the
log-likelihood that the observed density $Q(x)$ was drawn from the
model density $P(x)$. The $KLD$ is capable of describing also difference
between densities in faint tails. The $KLD$ itself can not be treated as a
distance directly since - even though it returns zero for identical
densities - it is not symmetric. We circumvent this problem by simply
symmetrising the $KLD$ and thus, we finally compute
\begin{equation}
KLD_{\mathrm{sym}}\left(P(x), Q(x)\right) = KLD\left(P,Q\right) +
KLD\left(Q,P\right)~.
\end{equation}

\subsection{Computation of distances}
The $KLD$ and the $BHA$ can not be computed analytically for two MoG and thus
must be approximated by performing the integration. Even though, analytical
approximations exist for the $KLD$ \citep[see][and references herein]{LowUp},
we encountered numerous difficulties when using them in practice, e.g., as
non-positive distances.
% and for the $L2$ norm the integral
% can be performed analytically.
% Unfortunately no such approximation exists for the BHA-distance. 
For this reason, we decided to perform the integration for all
distances numerically. 
%The Gauss-Legendre quadrature performs worse than the
%classical Riemann integral, while the adaptive quadrature yields better results
%but at the cost of significantly higher computation complexity. Finally, we
For our one-dimensional case, we found the following numerical integration to be
sufficient
\begin{equation}
\int_{-\infty}^{\infty} F(x)~dx \approx \Delta
\left(F\left(x_0\right)+F\left(x_0+\Delta\right)+\ldots+F\left(x_1\right)\right)~.
\end{equation}
The integration above is performed from $-\infty$ to $+\infty$. Here the
integral is numerically approximated and therefore a finite range must be
defined. The lower and the upper boundary are chosen very generously by 
integrating from $x_0=\mu\left(i\right)-5\sigma\left(i\right)$, 
with \mbox{$i = \underset{i \in
\mathrm{MoG}}{\mathrm{argmin}}{\mu\left(i\right)}$} to
\mbox{$x_1=\mu\left(i\right)+5\sigma\left(i\right)$}, 
with \mbox{$i = \underset{i \in
\mathrm{MoG}}{\mathrm{argmax}}{\mu\left(i\right)}$}.
In order to retain the same precision for all integrals we chose the integration
width $\Delta$ for all integrations to be the same. To be on the safe side, we set
$\Delta=0.001$ but when experimenting with this width it turned out that
$\Delta=0.005$ is sufficiently small to minimise computation time (scales
with $\Delta^{-2}$) without any loss in accuracy. 
Obviously, a good estimate for $\Delta$ is given by taking a
fraction of the typical standard deviation in the mixture of Gaussians, as
then the integration resolution is below the typical scale width of the
density. To prevent the integration from encountering ill defined
(that is, negative values in the square root or log) values we add a small
constant to each of the densities which does not yield any measurable impact on
the final classification.

\begin{figure}
\centering
\includegraphics[width=0.49\textwidth]{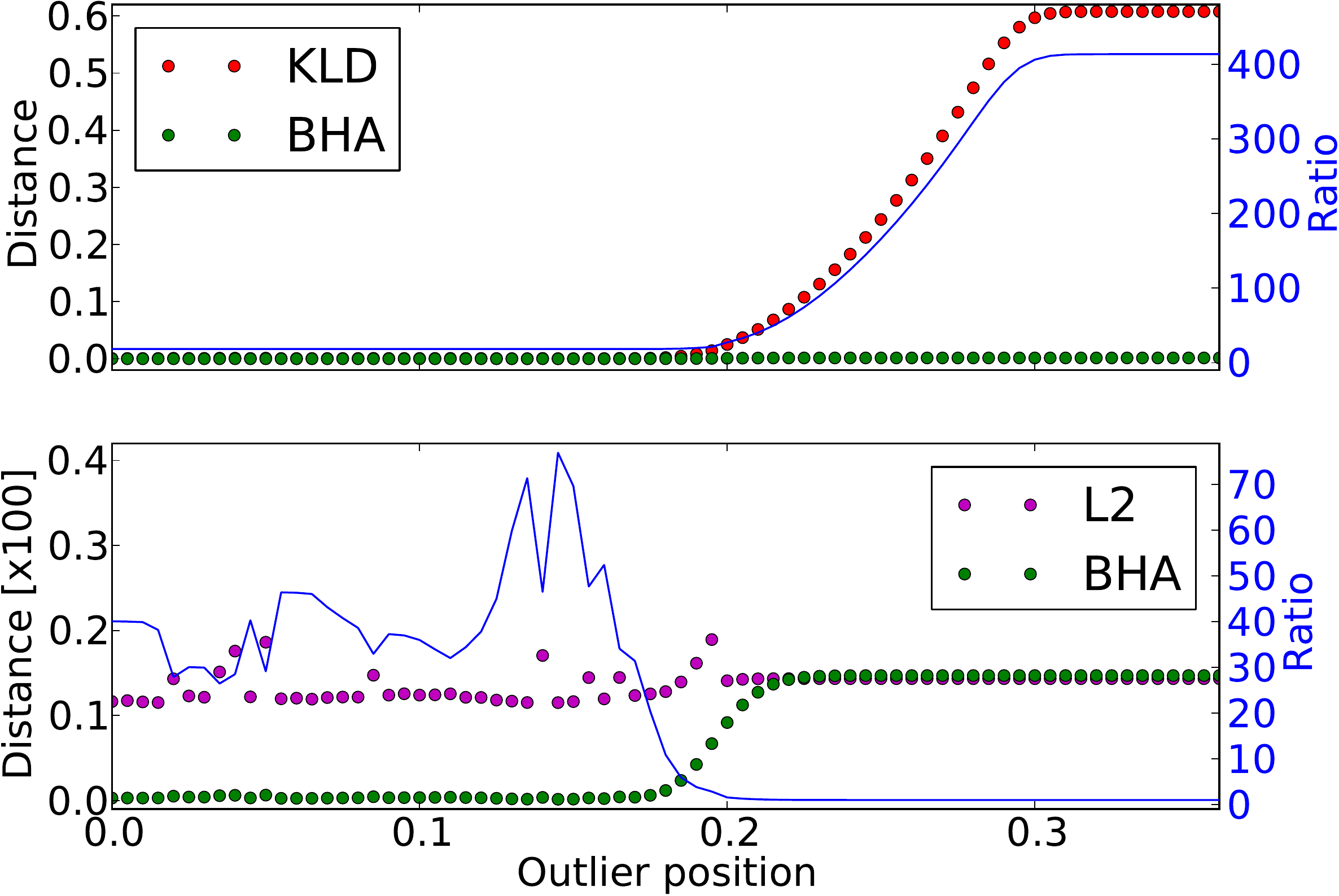}
\caption{The effect of an outlier added to the distribution in Figure
\ref{fig:hist2cont} is shown. The x-axis gives the magnitude of the injected
point with respect to the median, the left y-axis are the $KLD$/$BHA$ distance
in the upper plot and the $L2$/$BHA$ distance in the lower one. The right
y-axis denotes the ratio of the respective distance measures.
\label{fig:outlier}}
\end{figure}

In Figure \ref{fig:outlier} the impact of the injection
of a single outlier on the $L2$, $BHA$ and $KLD$ with respect to its injection
position (x-axis) is shown. Therefore a single measurement value with a typical
photometric error is inserted into the distribution from Figure
\ref{fig:hist2cont} and the distance to the undistorted distribution is
computed, respectively. It is evident, that the $KLD$ reacts way more heavily to
a single outlier, which will eventually also limits its use for classification
task, as shown in the results. Note, that in principle the $KLD$ distance would
diverge to infinity; the plateau is just encountered due to the added small
constant, mentioned above.

\subsection{Relation to features}
\label{sec:feat}
Our density representation directly relates to the 
features\footnote{To compute the features we use the python package provided at
\href{http://isadoranun.github.io/tsfeat/FeaturesDocumentation.html}
{http://isadoranun.github.io/tsfeat/FeaturesDocumentation.html}, also used in
\citet{2014ApJ...793...23N}.} used in \citet{2011ApJ...733...10R}; 
a detailed definition of all the used features is given in the \mbox{Appendix
\ref{sec:app}}.
As shown in Table \ref{featTable}, we can recover all but three
features directly from the density, except for \emph{StetsonK},
\emph{PercentAmplitude}, \emph{PercentDifferenceFluxPercentile (PDFP)}.
\emph{StetsonK} contains the discrete number of observations as one of its input
parameters, the latter two the absolute median value of the magnitudes. No
equivalent measures for those exist for our median-subtracted and normalised
densities.

To explain all the other features we use the common notion of moments of a
density
\begin{equation}
\sigma_n = \int_{-\infty} ^{+\infty}x^nP\left(x\right)dx
\end{equation}
with $\sigma_0=1$, $\sigma_1$ being the mean, $\sigma_2$ the standard
deviation and so on.
Another frequently used integral is the percentile, $x_f$, where the
density contains a certain fraction $f$, defined by
\begin{equation}
x_f: \int_{-\infty}^{x_f}P\left(x\right)dx=f.
\end{equation}
Additionally, the median absolute deviation (MAD) is defined as 
\begin{equation}
x_{MAD}: \int_{0}^{x_{MAD}}P\left(x-x_{0.5}\right) + P\left(x_{0.5}-x\right)dx=
0.5
\end{equation}
where $x_{0.5}$ is the median.
\begin{table}
\begin{tabular}{llrr}
Feature & Moment & Data & Model\tabularnewline
 & & Feature & Feature \tabularnewline
\hline
Amplitude & $A = 0.5 \cdot x_{0.95,0.05}$ & 0.150& 0.141 \tabularnewline
Beyond1Std & $1-\int_{\sigma_1-\sigma_2}^{\sigma_1+\sigma_2}P\left(x\right)dx$ &
0.446& 0.435 \tabularnewline 
FPRMid20$^{\star}$ & $x_{0.60,0.40}/x_{0.95,0.05}$&
0.325&0.309\tabularnewline 
FPRMid35$^{\star}$ & $x_{0.675,0.325}/x_{0.95,0.05}$ &
0.479& 0.468 \tabularnewline
FPRMid50$^{\star}$ & $x_{0.75.0.25}/x_{0.95,0.05}$& 0.625& 0.631
\tabularnewline 
FPRMid65$^{\star}$  & $x_{0.825,0.175}/x_{0.95,0.05}$ &
0.804& 0.789 \tabularnewline 
FPRMid80$^{\star}$  & $x_{0.90,0.10}/x_{0.95,0.05}$& 0.904&
0.899 \tabularnewline 
Skew & $\sigma_3/\sigma_2^3$ & -0.185& -0.182 \tabularnewline
SmallKurtosis & $\sigma_4/\sigma_2^4 - 3$& -1.365& -1.361 \tabularnewline
MAD & $x_{MAD}$
 & 0.083& 0.083 \tabularnewline 
 MedianBRP &
$\int_{x_{0.5}-A/5}^{x_{0.5}+A/5}P\left(x\right)dx$&
0.142& 0.126 \tabularnewline
PercentAmpl.$^{\dagger}$ & incl. median of LC& 0.013 & $-$ \tabularnewline
PDFP$^{\dagger}$ & incl. median of LC & 0.021  & $-$ \tabularnewline
StetsonK$^{\dagger}$ & incl. \# observations& 0.894  & $-$ \tabularnewline
%PercentAmplitude & & 0.013& 0.012 \tabularnewline
\tabularnewline
\cline{1-2}
\end{tabular}
{\tiny $^{\star}$FPR: FluxPercentileRatio, $x_{f,g} = x_{f} -
x_{g}$
\newline
$^{\dagger}$ Features without equivalent model description
}
\caption{\label{featTable}Computed
features from observations and from the density model with respective formulas
of an example light curve.}
\end{table}
One can see that most features can be expressed in terms of our $PDF$ and thus
the computed density contains most of the information encoded in the
features.

\subsection{Classification}
In this subsection we are describing the functionality and use of the different
classifiers applied in this work. The first two of those classifiers depend
actually on a distance matrix which is the direct outcome of our distance
measure. For the features the distance matrix is created by computing the
euclidean distance 
\begin{equation}
D\left(v, w\right) \equiv \sum_{n=1}^{N_{feat}}
\sqrt{\left(v_n-w_n\right)^2}
\end{equation}
between two feature vectors $v, w$. For the interested reader, more
details on the used classifiers can be found in \citet{TrevHast}. We use the
implementations provided in the python package 
scikit-learn\footnote{\href{http://scikit-learn.org/}{http://scikit-learn.org/}}.
To exclude effects originating from the preprocessing of the features, we also
classified the light curves with a min-max normalised version of the features.
In the following, the applied classifiers k nearest neighbors (kNN) and
the support vector machine (SVM) are explained in more detail.
\subsubsection{k nearest neighbours}
Once the distances between all light curves are computed, we can sort the matrix
for each candidate light curve and look at the types of the closest reference
light curves. The only free parameter is $k$, the number of neighbours chosen
per test light curve. Another degree of freedom can be introduced by weighting the
distances to the neighbours, e.g., decaying distance. In practice, we obtained
no significant gain and thus we use a classical majority vote. If the number of
objects of a certain class is equal for two (or more) different classes, a
random class out of those is assigned.

\subsubsection{Support vector machine}
A slightly better performance in classification can be reached if the distance
matrix is used as the kernel of a support vector machine (SVM). In this work, we
use the radial basis function (RBF) kernel which reads
\begin{equation}
K_{ij} = exp\left(-\frac{1}{\delta^2} D_{ij}\right)
\end{equation}
with $D$ being the distance matrix and $\delta$ the bandwidth.
As a consequence, low distances will have a kernel value close to unity and
distances significantly larger than $\delta$ will be close to zero.
We take the $\nu$-SVM as the kernel classifier. Two parameters have to be tuned
in a $\nu$-SVM, namely the kernel width $\delta$ and the width of the soft
margin $\nu$. The soft margin controls the fraction of mis-classifications in
the training of the classifier.

\subsubsection{Random forest}
Given the success in \citet{2011ApJ...733...10R}, we use the random forest (RF)
as a candidate classifier as well. RF extends the concept of a single decision
tree by using an ensemble of randomised decision trees.
Unfortunately, by its very nature, this classification method can only be used
on features. At each node of a tree the features are split such that the
information content (entropy) is maximised at each decision. The dominant free
parameter in a RF is the number of decision trees, which is the only one
considered in this work. 

\subsection{Performance and optimisation}
Each of the classifiers presented, has several free parameters to be optimised
but we stick for all the methods with the most important ones.
For the kNN comparison this parameter is the number of investigated $k$ nearest
objects, the $\nu$-SVM classifier has the tunable softening parameter $\nu$
and the kernel width $\delta$ and eventually the RF can be build up of $T$
number of trees.
While other parameters (e.g. tree depth in RF) might
have an impact on the classification quality, it is not the aim of
this work to investigate this possible gain with the choice of these
parameters. Also the process of feature selection is skipped and
throughout this work always all features defined in Table \ref{featTable} are
used. All the parameters are evaluated for each classifier and data set 
independently on a fixed grid and the respective value with the highest
accuracy is eventually chosen. A summary over the tuned parameters and
their respective search ranges, as well as all parameters that have not been
optimised, are shown in Table \ref{tab:paramGrids}.

\begin{table}
\begin{tabular}{lll}
classifier & parameter & range \tabularnewline
\hline
\multirow{2}{*}{kNN} & $\#$ neighbours $k$ & 1, 2, \ldots, 30\tabularnewline
 & weights & uniform (fixed) \tabularnewline

\rule{0pt}{3ex} & softening parameter $\nu$ & 0.01, \ldots, 1.00
 (adaptive)\tabularnewline
%$2^{-5}, 2^{-3}\ldots, 2^{13}$
\multirow{2}{*}{$\nu$-SVM} & kernel width $\delta$ & 0.01, \ldots, 100
 (adaptive)\tabularnewline 
 & kernel & RBF (fixed)\tabularnewline
 & kernel degree & 3 (fixed) \tabularnewline
 
 \rule{0pt}{3ex} & number of trees $T$ & 100, 200, \ldots, 1000
 \tabularnewline
 RF & split algorithm & gini (fixed) \tabularnewline
 & max. $\#$ of features & all (fixed) \tabularnewline 
\end{tabular}
\caption{\label{tab:paramGrids}
Overview over classifier parameters to be optimised}
\end{table}

We judge the performance of a classifier by computing the accuracy defined as
the mean fraction of correctly classified targets over a 10-fold
cross-validation; the uncertainty in accuracy is given by the standard
deviation.
In addition, we compute the confusion matrix of the best classifiers to
investigate possible caveats in the presence of multiple and unbalanced classes.
\section{Data}
\label{sec:data}
We conduct experiments with the different representations and classifiers
on two datasets. This has the advantage that we have two independent
measures for the predictive power of our method. In the first experiment, three 
classes are to be separated; in the second a more complex seven class
classification is performed. In fact, in the former dataset the classes
are defined more broadly (e.g., no distinction between different binary classes)
and thus it is expected that the classification accuracy will be higher than in
the latter case. It is the aim of this experiment to show, that our
classification algorithm can perform comparably well to state-of-the-art
classifiers for very broad and detailed classification tasks alike.

\subsection{OGLE}
The Optical Gravitational Lensing Experiment (OGLE,
\citealp{2008AcA....58..329U}) is a survey originally dedicated to the
search for microlensing events and dark matter. Therefore, stars of the
Magellanic clouds and the galactic bulge were monitored for the unique traces
of microlensing events. Consequently, millions of stars have been monitored,
delivering a rich database of variable stars. In our work, we use the dataset
used in \citet{2012ApJ...756...67W}\footnote{
\href{www.cs.tufts.edu/research/ml/index.php?op=data\_software}
{www.cs.tufts.edu/research/ml/index.php?op=data\_software}} 
where some RRLyrae, eclipsing binaries and cepheids in the
Magellanic clouds were extracted from the OGLE-II survey. 
The objects selected were known to be periodic before and thus their period
was known as well. In the publication,
the determination of the period is the main goal, but the database presents
a good test bed for classification as well, since a correctly determined
period favours also good classification results and thus the classification is
very reliable. The total number of objects is listed in Table \ref{OGLE_tab}.
Some of the files contain lines with invalid entries, that is a few lines with
a measurement error of zero, which have been removed.
\begin{table}
\begin{tabular}{llll}
Survey & VarType & Entities & $\langle$\#obs$\rangle$
\tabularnewline
\hline
& Cepheids & 3567 & 225 \tabularnewline
OGLE& Eclipsing binaries & 3929 & 330\tabularnewline
& RR Lyrae & 1431 & 323\tabularnewline
\hline
& MIRA & 2833 & 342 \tabularnewline
& ED & 2292 & 570 \tabularnewline
& RR Lyrae AB & 1345 & 412 \tabularnewline
ASAS& EC & 2765 & 524 \tabularnewline
& ESD & 893 &  547\tabularnewline
& DSCT & 566 & 492\tabularnewline
& DCEP-FU & 660 & 561\tabularnewline
\end{tabular}
\caption{Types of variables, number of entities and average
number of observations.\label{OGLE_tab}}
\end{table}
\subsection{ASAS}
The All Sky Automated Survey (ASAS, \citealp{1997AcA....47..467P}) is performed
with telescopes located on Hawaii and in Las Campanas and is lead by the Warsaw
university in Poland. The sky is observed in the I and V band with an initial
limit of 13\,mag (later extended to 14\,mag).  In 2005 the ASAS catalog of
variable stars (ACVS, \citealp{2005AcA....55..275P}) was published which is the
starting point for our experiment.
From the ACVS we extracted all objects with a unique classification which is
not miscellaneous. Subsequently, we removed all light curves having less than
50 observations and all classes with less than 500 members. 
A summary of the classes
used can be found in Table \ref{OGLE_tab}. For the classification we used the
magnitude ``Mag\_2'' (which corresponds  to a 4 pixel aperture) which is a
reasonably good measure of the brightness for fairly bright and faint stars.
Due to the extension to the faint end, the classes given could inherent
some false classifications itself, especially since also subclasses (e.g.,
detached and contact binaries) are annotated and hence, it is expected, the
class assignment in the given catalog is not as reliable as in the OGLE case.

\section{Results}
\label{sec:results}

\begin{figure}
\centering
\includegraphics[width=0.49\textwidth]{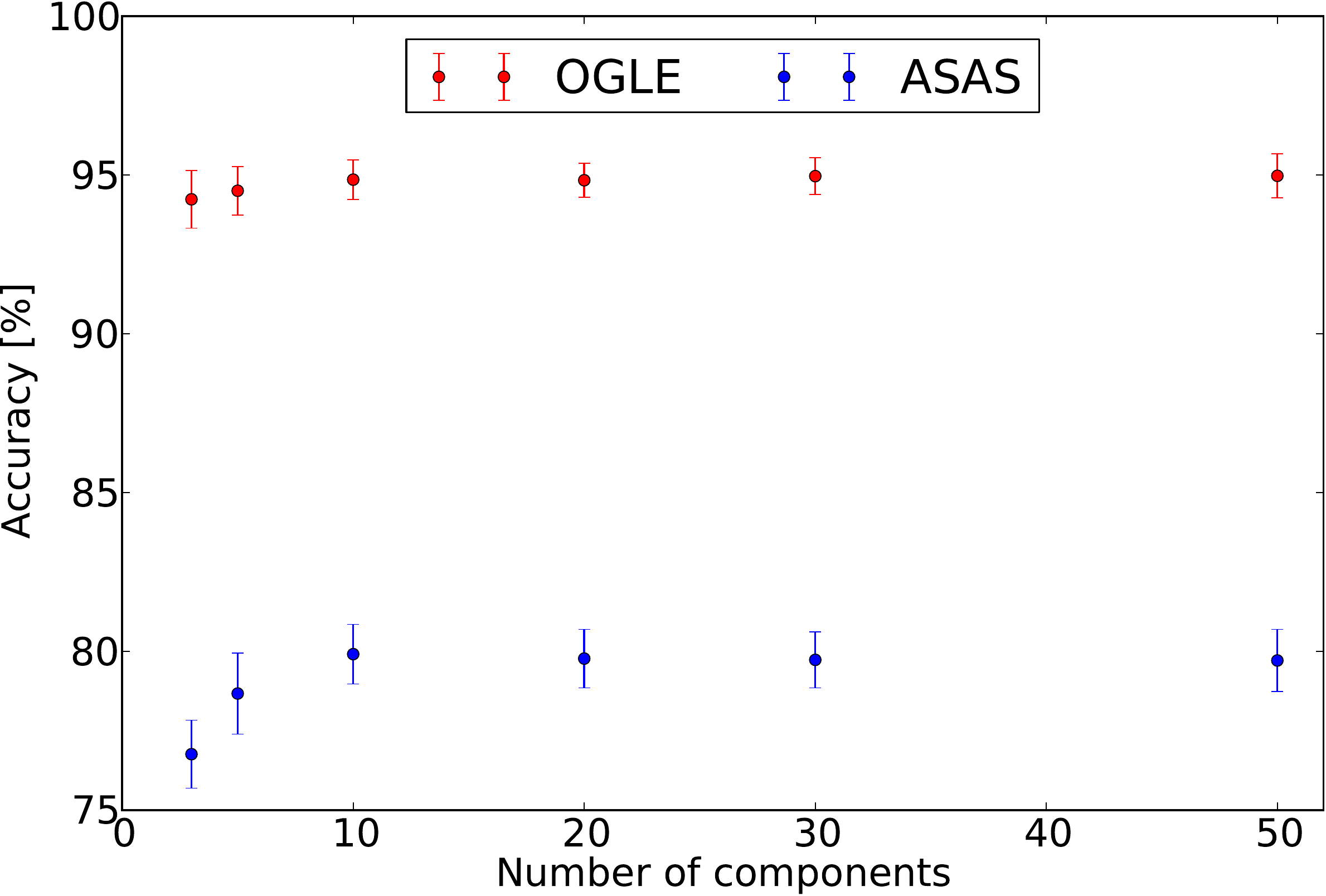}
\caption{Classification accuracy versus the number of Gaussian components using
the kNN classifier on both datasets. Accuracy is largely insensitive to the
exact choice of the number of Gaussian components.
\label{fig:ncomp}}
\end{figure}

As stated in the methodology, the impact of the reduction of the number of
Gaussians has to be quantified. In Figure \ref{fig:ncomp} we show empirically
that the impact on the final accuracy is only marginal, as long as the number
of components exceeds 10. For all conducted experiments we fix the number of
Gaussians to 20.

In Table \ref{tab:resTableOGLE} and \ref{tab:resTableASAS} the
results of the different experiments for the OGLE and ASAS data set are shown,
respectively. Since the $L2$-norm performs, independently of the chosen
classifier, always worse than the $BHA$ and $KLD$ metrics, we exclude it from
the discussion in the following. 

For the OGLE experiment we see that each method (feature and density
methods) performs comparably well within the typical deviation between the 10
cross-validation folds. It is worth noting, the RF, claimed to
be the best classifier in \citet{2011ApJ...733...10R}, does not perform any
better than the other classifiers. It is further interesting to see that the
feature-SVM is performing slightly better than the SVM based on the density
representation. As mentioned in Section \ref{sec:method}, three features exist
which cannot be described by the density-based approach. When removing those
respective features from the feature list, the accuracy of both feature-SVMs
drops by one per cent, indicating that the difference in accuracy does originate
from those.
The strength of the variation with respect to the median observed brightness
appears to bear some information about the type of variability. We elaborate
further on this issue in the discussion section.

\begin{table}
\centering
\begin{tabular}{llll}
 & kNN & $\nu$-SVM & RF \\
 \hline
Features & $95.09\pm0.74$ & $96.86\pm0.52$  & $95.61\pm0.82$\\
(raw) & $k$=8 & $\nu=$0.04, $\delta=$0.31 & $T$=500\\
\\
Features & $95.51\pm0.81$ & $96.88\pm0.67$ &
$95.59\pm0.83$\\
(norm.)& $k$=10 & $\nu=$0.06, $\delta=$0.08 & $T$=500\\
\\
\multirow{2}{*}{$L2$} & $93.44\pm0.88$ & $95.92\pm0.68$ & $-$\\
& $k$=3 & $\nu=$0.06, $\delta=$0.69& $-$\\
\\
\multirow{2}{*}{$KLD$} & $95.14\pm0.70$ & $95.51\pm0.94$ & $-$\\
& $k$=5 & $\nu=$0.14, $\delta=$0.33& $-$\\
\\
\multirow{2}{*}{$BHA$} & $94.84\pm0.83$ & $96.01\pm0.71$ &
$-$\\
& $k$=7 & $\nu=$0.08, $\delta=$0.14& $-$\\
\end{tabular}
\caption{Results for the optimal classifiers for the 3 class classification of
OGLE data. The performance is the average fraction of correctly classified
objects in a 10-fold cross-validation with the standard deviation of this
performance being the error (all given in per cent).
\label{tab:resTableOGLE}}
\end{table}

\begin{table}
\centering
\begin{tabular}{llll}
 & kNN & $\nu$-SVM & RF \\
 \hline
Features & $74.22\pm1.24$ & $78.02\pm0.68$ &
$79.98\pm1.16$\\
(raw) & $k$=11 & $\nu=$0.19, $\delta=$0.53 & $T$=400\\
\\
Features & $77.60\pm0.76$ & $80.47\pm1.21$ &
$79.99\pm1.55$\\
(norm.)& $k$=17 & $\nu=$0.17, $\delta=$0.10 & $T$=400\\
\\
\multirow{2}{*}{$L2$} & $79.57\pm0.80$ & $82.08\pm0.89$ &
$-$\\
& $k$=19 & $\nu=$0.01, $\delta=$0.56& $-$\\
\\
\multirow{2}{*}{$KLD$} & $78.96\pm1.87$ & $75.56\pm0.94$ &
$-$\\
& $k$=23 & $\nu=$0.26, $\delta=$0.34& $-$\\
\\
\multirow{2}{*}{$BHA$} & $79.73\pm0.83$ & $81.11\pm0.90$ &
$-$\\
& $k$=29 & $\nu=$0.20, $\delta=$0.14& $-$\\
\end{tabular}
\caption{Results for the optimal classifiers for the 7 class classification of 
ASAS data. The performance is the average fraction of correctly classified
objects in a 10-fold cross-validation with the standard deviation of this
performance being the error (all given in per cent).
\label{tab:resTableASAS}}
\end{table}

\begin{figure}
\centering
\includegraphics[width=0.49\textwidth]{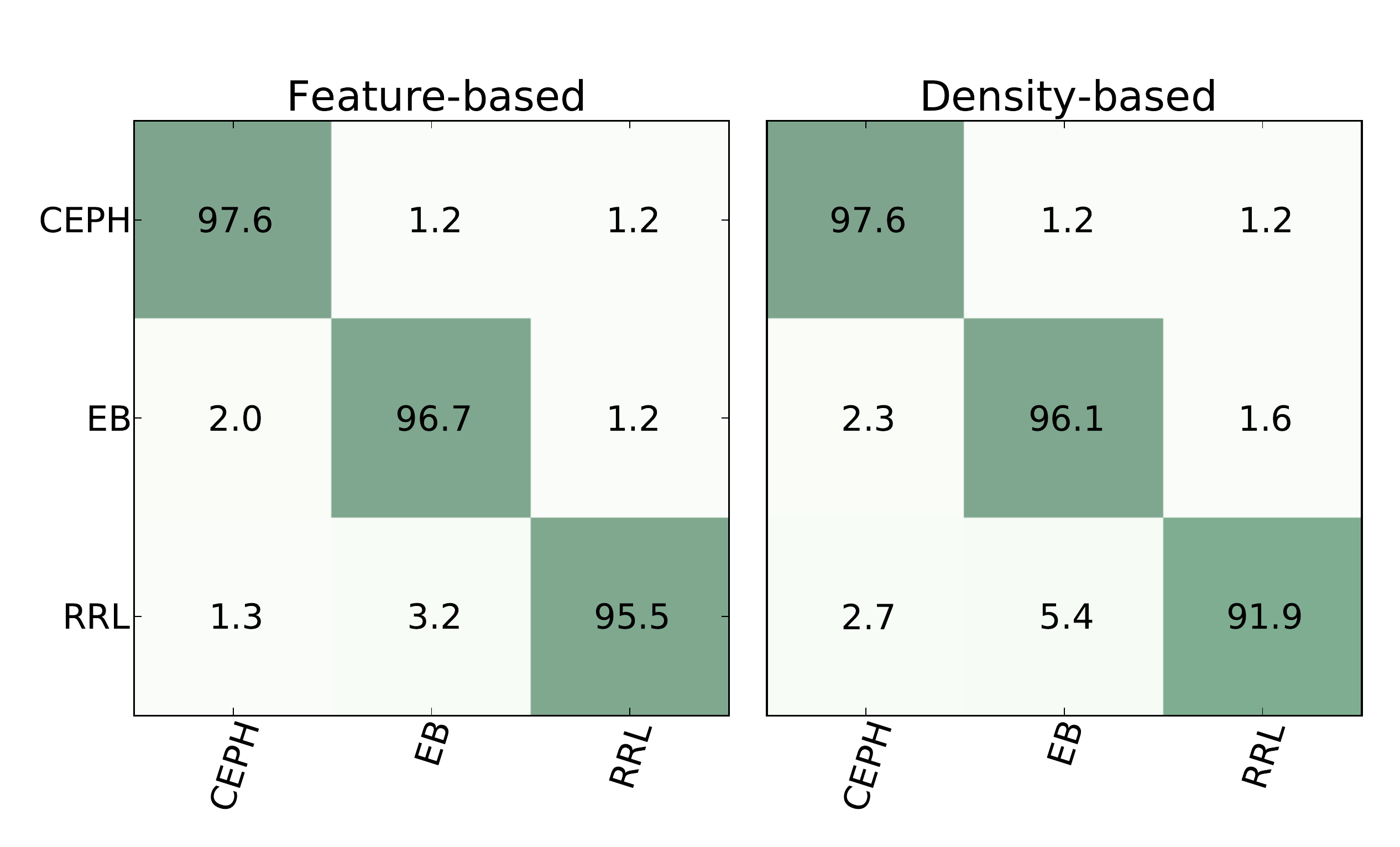}
\caption{The accuracies (given in per cent) of the feature based
(left) and density ($BHA$-metric) based $\nu$-SVM classifier are shown for the
OGLE dataset.
The x-axis shows the labels according to the classifiers, the y-axis the given
ones; the colour scale stands for the respective accuracy; from zero (red) to 
hundred (green) per cent.
\label{conf_ogle}}
\end{figure}

\begin{figure}
\centering
\includegraphics[width=0.49\textwidth]{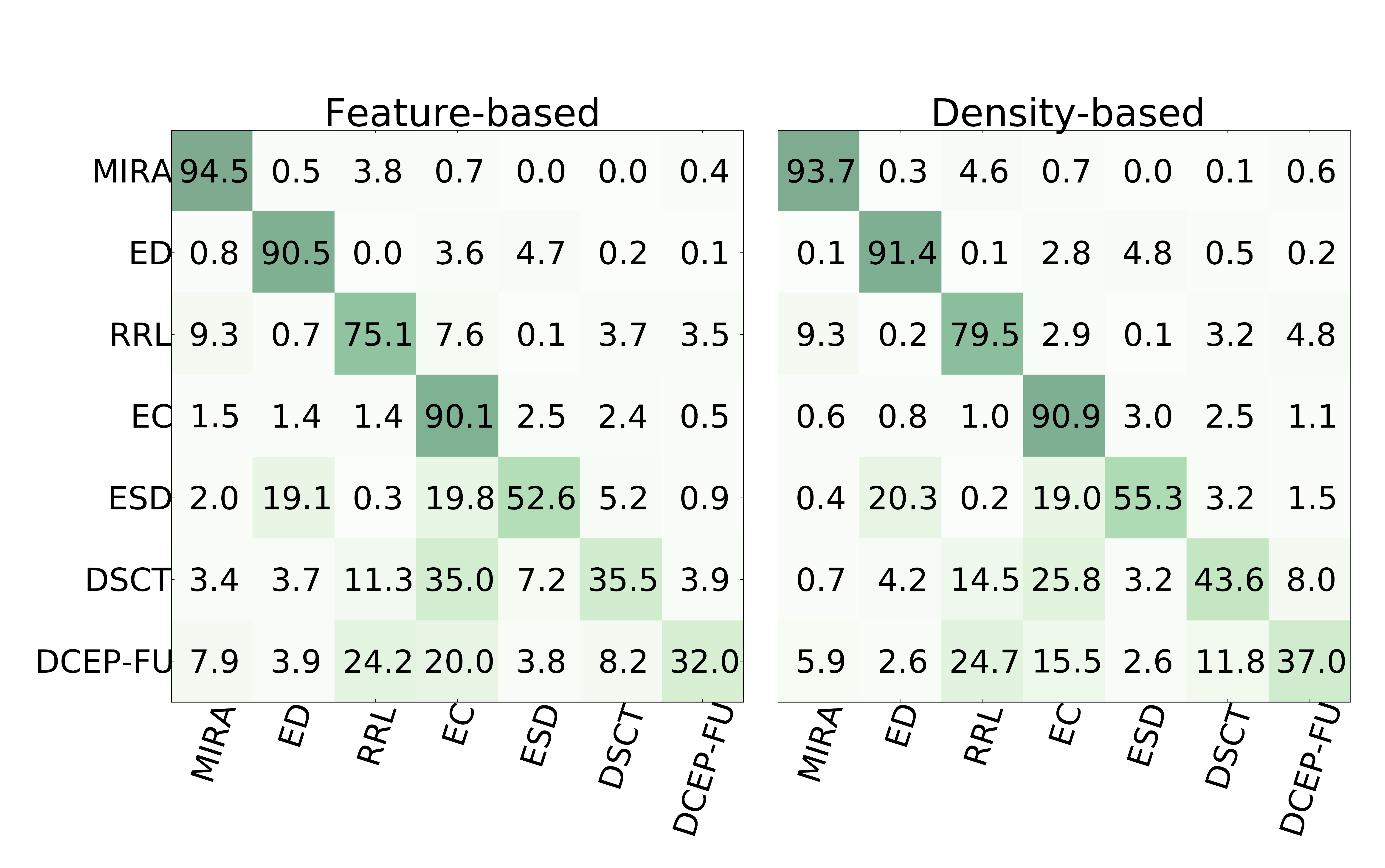}
\caption{The accuracies (given in per cent) of the feature based
(left) and density ($L2$-metric) based $\nu$-SVM classifier are shown for the
ASAS dataset.
The x-axis shows the labels according to the classifiers, the y-axis the 
given ones; the colour scale
stands for the respective accuracy; from zero (red) to hundred (green) per cent.
\label{conf}}
\end{figure}
That the impact of those median-based features is anyway not too high is
supported by the results of the seven class ASAS classification. It becomes
apparent that the more generic definition of the density enhances the
accuracy in contrast to all the feature-based classifiers. 
The confusion matrices of the best classifiers from the density and feature
based classification are shown in \mbox{Figure \ref{conf_ogle}, \ref{conf}}. It
can be seen that classes with more members achieve a higher accuracy which is
expected due to the higher number of training objects. Otherwise, no
significant biases in any direction between the two different classification
approaches can be detected.
While the gain in
accuracy is again only marginal, it can be shown that the same
quality is only reached if the three features, not describable by the densities,
are included. Else the classification rates of the feature-based SVMs drop again
by one per cent. Apart from this, it can be observed that the classification
quality of the density-based classifiers depends quite strongly on the choice
of the distance metric. The $KLD$ does perform in three of the four experiments
worse than the $BHA$ which supports the statement that the $BHA$ distance is a
good distance measure for classification tasks. On the other hand, one should
realise that the choice of the metric, that is the distance between two given
feature vectors, is in principal also a free methodological factor in the
classification problem. Apart from the standard euclidean distance and the
Mahalanobis distance no other measures have been investigated in the
literature.

%The computational
%complexity of the random forest classifier is given by $\propto
%\mathcal{O}\left(dMN\mathrm{log}_{10}N\right)$, where d is the dimensionality
%(number of features), M the number of trees and N the number of entities. The
%scaling complexity for a structured kNN search as applicable in our case is of
%the order of $\propto \mathcal{O}\left(dN\mathrm{log}_{10}N\right)$ where d is
%the number of the quasi-continuous grid points.

\section{Discussion}
\label{sec:discussion}
In this work, we present a generalisation of static features for the
classification of time series. In contrast to previous work, we do not rely
on describing static densities with a set of features but use the
densities themselves to measure the similarity between two light curves. By
doing so, we can reduce the number of degrees of freedom in the methodology
from four (preprocessing $-$ feature selection $-$ choice of metric $-$ choice of
classifier) to two (choice of metric $-$ choice of
classifier). This allows us to skip the step of feature selection.
The proposed approach follows first principles by simply
assuming a model for representing the data; once a metric is chosen,
classification in a kernel setting follows naturally.
The strong point of the newly proposed representation is the
fact that it captures all the information present in the data (including
measurement errors) and makes it available to the classifier.

As highlighted in the results, the choice of
the metric used in the density representation plays an important role.
A priori, we are not aware of any natural choice of a metric. We have
shown in our experiments that the $BHA$ and $L2$ distance are
performing very well in terms of accuracy. In principle, other (or combinations
of) metrics might exist that are more suited for a given classification problem. 

Our approach presents a different way of performing classification. Therefore,
it provides independent evidence that the widely used features are indeed well
chosen for the classification problems considered so far. However, it is unclear
how well the chosen features generalise to other classification problems. On the
other hand, the density representation is formulated generically and encodes all
information available in the data. Additionally, the proposed method naturally
encodes also uncertainty in the measurements, which is not taken into account in
the feature-based approaches so far. As a consequence, it is now possible to
learn a classification on data of one survey that contains small (large)
measurement uncertainty, and predict on data of another survey with large
(small) photometric error. While this is problematic for feature-based
approaches, it is automatically taken care of in the density representation.

\begin{figure}
\centering
\includegraphics[width=0.49\textwidth]{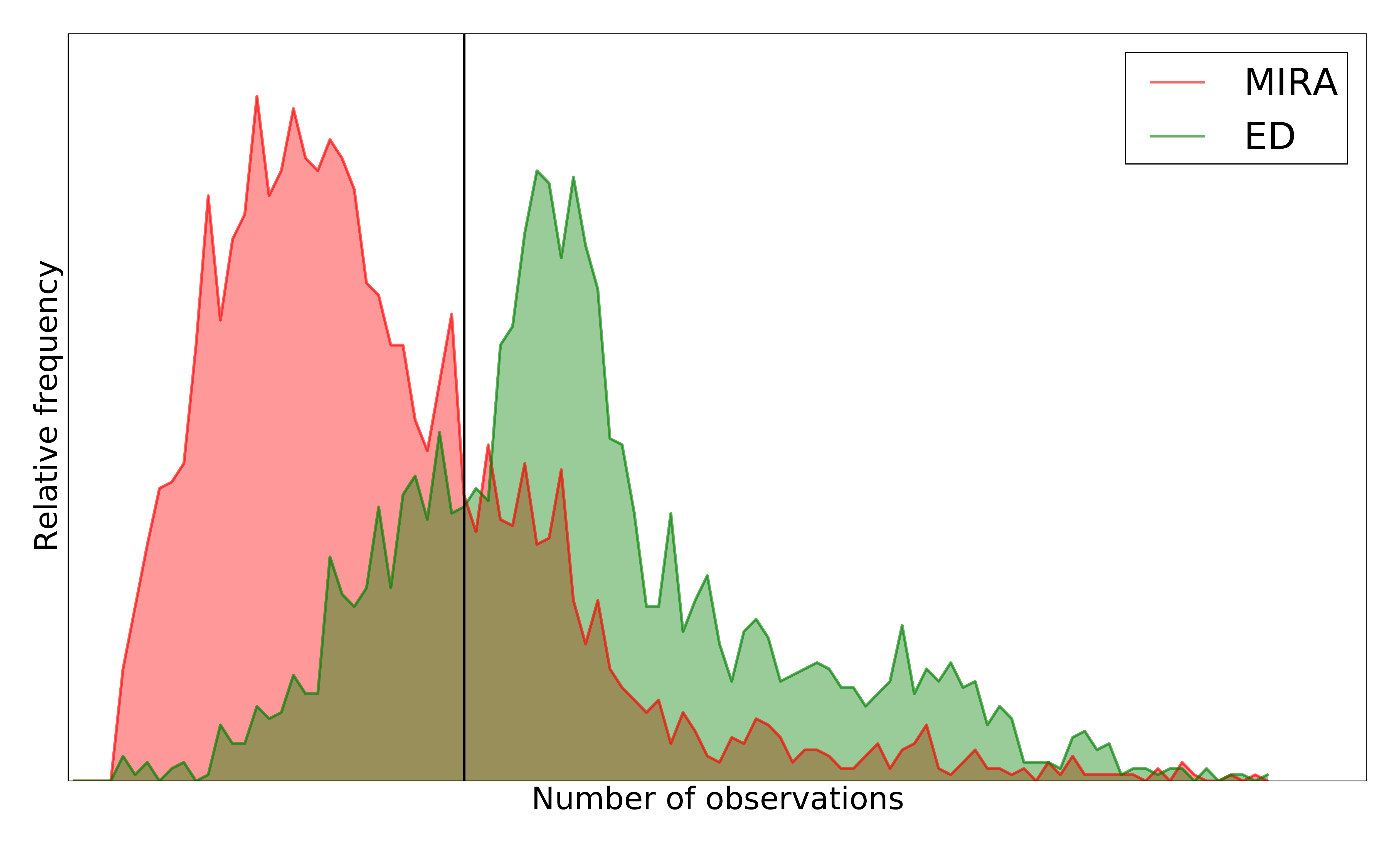}
\caption{A histogram over the number of observations for the MIRA and detached
binaries classes in the ASAS survey are shown. The number of observations
clearly correlates with the class label: if a bisectional line is introduced at
427 observations, a classification rate of 75.1\% 
can be reached.
\label{bias}}
\end{figure}

We have shown that the feature- and density-based approaches perform comparably
well in terms of accuracy for the given datasets. As aforementioned, there are
three features that cannot be derived from the density representation which
appear to increase the classification accuracy. In particular, the
\emph{StetsonK}
 value depends directly on the number of observations in a light
curve, and for this reason it cannot be derived 
from our representation. It is questionable why
the number of observations should be a defining property of a class. The only
reason why it contributes to the performance is because certain classes are
apparently observed more often than other ones (see Table
\ref{OGLE_tab}), and not because it is an inherent physical property. In Figure
\ref{bias} we show that it is possible to classify ASAS light curves into MIRA
and detached binaries with a 75\% accuracy solely using the number of
observations that happened to be recorded. 
The brightness of stars that vary over a wide range of magnitudes, such as
MIRA, will frequently drop below the survey-specific detection limit.
Hence, faint observations will not be recorded in the database. This raises the
following problem: absent recordings are ambiguous because it
is not evident whether the source was too faint to be detected or simply not
observed. As a consequence, the number of observations, and thus
\emph{StetsonK}, hints to the variability type of a star within one survey.
However, this feature is survey-dependent as surveys differ in database
structure (e.g., some give upper limits) and detection limits, and thus does
not generalise.
If non-detections are not treated accordingly, 
the definition and use of the \emph{StetsonK} value can
cause dramatic bias on the classification, especially when knowledge is
transferred between different surveys, as done, e.g.
in \citet{2010ApJ...713L.204B}. Similarly, the \emph{PercentAmplitude} and the
\emph{PDFP} directly depend on the apparent magnitude of the respective object,
which is also not an inherent property of a class. Conclusively, the only
reason why these three features contribute to the accuracy is because of the
presence of a (or several) bias in the observations and not because they
capture physical characteristics of the data.
We do not state that the features in question are useless for classification
(indeed they increase the accuracy), but argue that they do not generalise and
are therefore not useful for knowledge transfer between surveys with different 
observational bias. It should be considered, to redefine these features
accordingly, such that they do not rely on the observation strategy of a
survey.

In summary, the proposed method (a) introduces a more general notion of distance
between light curves in contrast to static features, (b) naturally incorporates
measurement errors, (c) performs equally well as state of the art
feature-based classifications and (d) 
yields an independent measurement of the accuracy as compared to feature-based
classification.
 
As a future prospect, the density-based representation could be useful in
unsupervised settings where the notion of distance is more critical in the
absence of labels which are the driving force in a classification task. Feature
sets that have been optimised for classification do not necessarily provide a
good similarity measure. In subsequent work, we will investigate whether the
proposed notion of distance naturally distinguishes between the different
variability types. Additionally, we advocate that besides
static features also temporal information should be incorporated in a similar
vein. However, the design of such a time-dependent representation remains an
open question.

\section*{Acknowledgments}
SDK would like to thank the Klaus Tschira Foundation for their financial
support. 
The authors would like to thank the anonymous referee for his
interest and his very helpful suggestions.
% Since our best classification method
% (SVM based on a Bhattacharyya metric) outperforms the random forest and the
% feature based SVM only marginally, we can validate that the chosen features are
% well suited for this classification problem.
% 
% As highlighted in the results, this statement does not hold
% generally, as not all static features can be represented in our current
% representation of time series.  
% Currently we are investigating the possibility to expand this more generic
% method to also include static Amplitude(features which are based on the
% median, as well as quasi-static features. At the moment it is not clear, whether this methodology
% can be used also to describe light curves (and their full time behavior) in
% general, but even if ``only'' the question of the impact and use of
% (quasi-)static features can be resolved, a great leap forward in
% the\ref{sec:feat} understanding of irregulary sampled time series has been made.
%\begin{thebibliography}{99}
\bibliography{library.bib}
\bibliographystyle{mn2e}
% \input{}\href{http://isadoranun.github.io/tsfeat/FeaturesDocumentation.html}
%{http://isadoranun.github.io/tsfeat/FeaturesDocumentation.html}
%\end{thebibliography}
\appendix
\section[]{Detailed description of features}
\label{sec:app}
In the following, we give a detailed description of the static features used in
\citet{2011ApJ...733...10R}.
The computation of the software is done using the Python FATS package, available under
\href{https://pypi.python.org/pypi/FATS}{https://pypi.python.org/pypi/FATS}. The
error in the definition of the StetsonK value in older versions was corrected
manually.
\newline\newline
\noindent\emph{Amplitude} Absolute difference between highest and lowest
magnitude.
\newline\newline
\noindent\emph{Beyond1Std} Fraction of photometric points that lie beyond one standard
deviation with respect to the (with photometric errors) weighted mean.
\newline\newline
\noindent\emph{FluxPercentileRatio (FPR)} Relative difference of flux percentiles with
respect to the 95 to 5 percentile difference. The number after the FPR gives the
width of the percentile, always centered on 50, e.g., $FPR20 = \frac{F_{60} -
F_{40}}{F_{95} -F_{5}}$.
 \newline\newline
\noindent\emph{Skew} The skew of the distribution of magnitudes. 
\newline\newline
\noindent\emph{SmallKurtosis} Kurtosis of the magnitudes for small samples.
\newline\newline
\noindent\emph{Median absolute deviation (MAD)} Median deviation of the absolute
deviation from the median.
\newline\newline
\noindent\emph{Median buffer range percentage (MedianBRP)} Fraction of data points
lying within one tenth of the amplitude around the median. 
\newline\newline
\noindent\emph{PercentAmplitude} Largest absolute difference from the median magnitude,
divided by the median magnitude itself.
\newline\newline
\noindent\emph{PercentDifferenceFluxPercentile (PDFP)} The 95 to 5 flux percentile
difference, divided by the median of the flux.
\newline\newline
\noindent\emph{StetsonK} More robust measure of the kurtosis, as defined in 
\citet{1996PASP..108..851S}.
\label{lastpage}

\end{document}